\documentclass[
  ,final            
  ,numberedheadings 
  ] {aipproc}

\layoutstyle{6x9}

\begin{document}

\title{Interstellar dust: what is it, how does it evolve, and what are its observational consequences?}

\author{Eli Dwek}{
  address={Laboratory for Astronomy and Solar Physics\\ NASA Goddard Space Flight Center \\ Greenbelt, MD 20771, U.S.A.\\ e-mail: eli.dwek@nasa.gov}
}

\begin{abstract}
The presence of dust in the interstellar medium of galaxies dramatically affects their spectral appearance, and the interpretation of their underlying physical properties.  Consequently, the diagnostic of galaxy spectra depends crucially on our understanding of the nature and properties of these dust particles. Models of interstellar dust particles must be able to reproduce a basic set of observational constraints, including the general interstellar extinction and infrared emission observed in the diffuse interstellar medium (ISM). Recent analysis of the solar spectrum have resulted in a convergence between the solar and B star abundances. This development, and the steadiness in the solar abundance determination of the primary refractory elements Mg, Si, and Fe, strongly suggest that any viable dust model must also obey interstellar abundances constraints. Fifteen dust models that differ in composition and size distribution and that simultaneously satisfy the local extinction, infrared emission, and abundances constraints have been shown to exist. This multitude of viable dust models provides us with an increased flexibility in understanding dust evolution and the many variations in dust properties in different phases of the ISM and stellar environments. 

The evolution of stars, the primary sources of dust, and the general ISM, in which  the dust is processed, change the abundance of dust as a function of time. In particular, the delayed injection of dust from low mass stars, which are the primary sources of carbon dust, into the ISM gives rise to a changing silicate-to-carbon dust mass ratio, which will affect the UV-visual extinction in galaxies. If PAHs are only produced in AGB stars, the delayed injection of these molecules into the ISM may be in part  responsible for the absence of PAH features in young star forming regions, or for the existence of a metallicity threshold below which PAHs have not yet formed. 

On a cosmological scale, an epoch of rapid and efficient dust formation is needed to account for the presence of dust-enshrouded star forming regions at high redshift. We show that the first opacity is probably produce by SN-condensed carbon dust that formed in less than $\sim$ 100~Myr after the onset of galaxy formation.
\end{abstract}

\maketitle


\section{Introduction}
Dust is present in almost every astrophysical environment, ranging from circumstellar shells and disks
to spiral, elliptical, starburst, and active galaxies, and to pre$-$galactic objects such
as QSO absorption$-$line and damped Ly$\alpha$ systems. The abundance and composition
of the dust in galaxies affect the galaxies' spectral appearance, and influence
the determination of their underlying physical properties, such as their star formation
rate, metallicity, and attenuation properties. 
Understanding the properties of interstellar dust particles is therefore essential for the interpretation of galactic spectra. 

In this manuscript I will briefly review what constitutes an interstellar dust model, list the observational constraints on such models, and briefly describe viable interstellar dust models that satisfy these constraints in the local interstellar medium (ISM). Special emphasis will be placed on the interstellar abundance constraints, which until recently, have not been explicitly included in dust models. 



Interstellar dust exhibits spatial and temporal variations, and I will briefly review the ingredients in constructing models for the evolution of dust, stressing the current uncertainties in the yield of dust from supernovae and AGB stars. Finally, I will describe the effect of dust evolution on the spectral energy distribution of galaxies, and, using a very simple criterion, present a simple estimate of the redshift when galaxies first become opaque.

More detailed information on observational aspects of interstellar dust and the physics of dust can be found in the recent review article by Draine \cite{d04}, in the workshop on "Solid Interstellar matter: The ISO Revolution" \cite{djj} and the conference on "Astrophysics of Dust" \cite{wcd}, in the books by Whittet \cite{w03} and Kr\"ugel \cite{k03}, and the recent issue of The Astrophysical Journal Supplement Series (volume 154) dedicated to the first observations with the {\it Spitzer} Space Telescope.

\section{Interstellar dust models}
\subsection{What constitutes an interstellar dust model?}
 An interstellar dust model is completely characterized by the abundance of the different elements locked up in the dust, and by the composition, morphology, and size distribution of its individual dust particles. This seemingly simple definition hides the complexities involved in deriving such a dust model.
 
 First and foremost, any dust model must specify the total mass of the different refractory elements that are locked up in the solid phase of the ISM. These elements can form many different solid or molecular compounds with different optical and physical properties. In addition, the morphology of the dust particles whether spherical, ellipsoidal, cylindrical, platelike, or amorphous has an important effect on these properties. Finally, the size distribution of these dust particles will determine their collective properties and interactions with the ambient gas and radiation field. These interactions  play a major role in the radiative appearance of galaxies, and in the thermal and chemical balance of their interstellar medium.    

\subsection{Observational constraints in the local ISM}
Ideally, a viable interstellar dust model  should fit all observational constraints  arising primarily from the interactions of the dust with the incident radiation field or the ambient gas. These include:
\begin{enumerate} 
\item the extinction, obscuration, and reddening of starlight; 
\item the infrared emission from circumstellar shells and different phases of the ISM (diffuse H~I, H~II regions, photodissociation regions or PDRs, and molecular clouds);
\item the elemental depletion pattern and interstellar abundances constraints;
\item the extended red emission seen in various nebulae;
\item the presence of X-ray, UV, and visual halos around time-variable sources (X-ray binaries, novae, and supernovae);
\item the presence of fine structure in the X-ray absorption edges in the  spectra of X-ray sources; 
\item the reflection and polarization of starlight;
\item the microwave emission, presumably from spinning dust;
\item the presence of interstellar dust and isotopic anomalies in meteorites and the solar system;
\item the production of photoelectrons required to heat neutral photodissociation regions (PDRs); and
\item the infrared emission from X-ray emitting plasmas.
\end{enumerate}

It is unreasonable to require that a single dust model simultaneously fit all these observational constraints since they vary in different astrophysical environment, reflecting the regional changes in dust properties. 

However, a viable interstellar dust model should be derived by {\it simultaneously} fitting at least a basic set of observational constraints. Also, it must consist of particles with realistic optical, physical, and chemical properties, and require no more than the ISM abundance of any given element to be locked up in the dust. In practice, most interstellar dust models have been constructed by deriving the abundances and size distributions of some well studied  solids, such as graphite or silicates, using select observations such as the average interstellar extinction, the polarization, or the diffuse infrared emission as constraints, and then checked the model for consistency with other observational constraints such as the wavelength dependent albedo and interstellar abundances.

Simultaneous fits to the average interstellar extinction curve and the infrared (IR) emission from the diffuse ISM have given rise to a standard interstellar dust model consisting of bare, spherical graphite and silicate particles and a population of polycyclic aromatic hydrocarbons (PAHs) \cite{ld01}. This standard model can account for the observed 2175 \AA \ bump in the UV extinction curve and the far-UV rise in extinction attributed to graphite and PAHs; for the mid-IR emission features at 3.3, 6.2, 7.7, 8.6, and 11.3 $\mu$m, most commonly associated with PAHs; and for the general continuous IR emission from the diffuse ISM, attributed to submicron size silicate and graphite grains. The choice of graphite and PAHs was mainly motivated by the UV extinction bump and the mid-IR emission features. Silicate particles were included to account for the presence of the 9.7 and 18~$\mu$m absorption features seen in a variety of astrophysical objects and Galactic lines of sights.  Using this model as a benchmark, regional variations in the observational manifestations of the dust can then be attributed to local deviations from this standard model. For example: the lack of the mid-IR emission features from inside H~II regions can be attributed to the depletion of PAHs in these regions; the flattening of the extinction curve at UV energies to a deficiency in PAHs and very small dust particles; and the presence of various absorption features in the spectra of some astronomical sources to the precipitation of ices onto the dust in these objects.  Brief reviews on the history of the development of dust models were presented by \cite{d03, d04, dw04a}. 

\subsection{Interstellar dust models with cosmic abundances constraints}
An important advance in the construction of interstellar dust models was made by Zubko, Dwek, \& Arendt \cite{zda} (hereafter  ZDA). The ZDA approach differs from previous dust modeling efforts in two important ways: first, it includes, in addition to the average interstellar extinction and diffuse IR emission, cosmic abundances as an {\it explicit} constraint  on the models; and second,  it solves the problem of simultaneously fitting these three observational constraints by an  inversion method called the method of regularization. Uncertainties in the data are propagated into uncertainties in the derived grain size distribution. 

Interstellar abundances were previously not used as explicit constraints in dust models because of the large discrepancies between abundances inferred from solar, F and G stars, and B stars measurements (see \cite{s04} for a recent review). In particular, B star carbon abundances were found to be significantly different from solar abundance measurements given by, for example, Holweger \cite{h01}. This discrepancy precipitated an interstellar carbon "crisis" \cite{sw95, km96}, since standard interstellar dust models by Mathis, Rumpl, \& Nordsieck \cite{mrn} or Draine \& Lee \cite{dl84} required more carbon to be locked up in dust than available in the ISM.  Mathis \cite{m76} attempted to solve the crisis by suggesting that most of the interstellar carbon dust is in the form of amorphous fluffy dust.  However, Dwek \cite{dw97b} showed that the Mathis model failed to include the amount of carbon needed to produce the PAH features, and that the fluffy carbon particles produced an excess of far-IR emission over that detected by the {\it COBE} satellite from the diffuse ISM \cite{dw97a}. 

A very recent analysis of solar absorption lines by Asplund, Grevesse, \& Sauval \cite{ags}, which included the application of a time-dependent 3D hydrodynamical model for the solar atmosphere, has led to a dramatic revision of the abundance of carbon in the sun. The revised carbon and oxygen abundances are now in much better agreement with local ISM \cite{an03}, and with the B star abundances, which are commonly believed to represent those of the present day ISM. 
     
ZDA considered  five different dust compositions as potential model ingredients: (1) PAHs; (2) graphite; (3) hydrogenated amorphous
carbon of type ACH2; (4) silicates (MgSiFeO$_4$); and (5) composite particles
containing different proportions of silicates, organic refractory material (C$_8$H$_8$O$_4$N), water ice (H$_2$O), and voids. 
These different compositions were used to create five different classes of dust models:
\begin{itemize}
\item {\bf The first class} consists of PAHs, and bare graphite and
silicate grains, and is identical to the carbonaceous/silicate model recently
proposed by \cite{ld01}. \\
\item  {\bf The second class} of models contains composite particles in addition
to PAHs, bare graphite and silicate grains. \\
\item  {\bf The third and fourth classes} of models comprise the first and second
classes, respectively, in which the graphite particles are completely replaced by amorphous
carbon grains. \\
\item  {\bf In the fifth class} of models the only carbon is in PAHs and in the organic refractory material in composite grains. That is, the model comprises only PAHs, bare silicate, and composite particles. 
\end{itemize}

To accommodate the uncertainties in the ISM abundances, ZDA considered three different ISM abundance determinations: solar, B-star, and F-G star abundances, as constraints for the dust models. 
The method of regularization proved to be a robust method for deriving the grain size distribution and abundances (the two unknowns) for the different classes of dust models. The results show that there are many classes of interstellar dust models that provide good simultaneous fits to the far-UV to near-IR extinction, thermal IR emission, and elemental abundances constraints. The models can be grouped into two major categories: BARE and COMP models. The latter are distinguished from the former by the fact that they contain a population of composite dust particles which generally have larger radii than bare particles. 

\subsection{Results and astrophysical implications}

Table 1 compares the abundances of refractory elements  in two ZDA and the Li \& Draine \cite{ld01} dust models to the constraints imposed from abundance determinations in stars and the ISM. Abundances are normalized to 10$^6$ hydrogen atoms or parts per million (ppm). The abundances in the dust were derived by subtracting the observed gas phase abundances from the respective solar, F and G stars, and B star abundances. Note that  the F and G star abundances have larger uncertainties in their O, Mg, and Si determinations than their solar and B stars counterparts. However, all abundances are consistent within the 1$\sigma$ uncertainties in their determinations.


\begin{table}
\begin{tabular}{lllllll}
\hline
\tablehead{1}{l}{b}{ }
&\tablehead{1}{l}{b}{reference}
  & \tablehead{1}{l}{b}{C}
  & \tablehead{1}{l}{b}{O}
  & \tablehead{1}{l}{b}{Mg}
    & \tablehead{1}{l}{b}{Si}
  & \tablehead{1}{l}{b}{Fe}   \\
\hline
\bf{Total}& Solar \tablenote{Asplund, Grevesse, \& Sauval \cite{ags}} & 245$\pm$30 & 457$\pm$56 & 34$\pm$8 & 32$\pm$3    & 28$\pm$3\\
 & F \& G stars \tablenote{Sofia \& Meyer \cite{sm01}} &358$\pm$82 & 445$\pm$156 & 43$\pm$17 & 40$\pm$13    & 28$\pm$8\\
& B stars \tablenote{Sofia \& Meyer \cite{sm01}} &190$\pm$77 & 350$\pm$133 & 23$\pm$7 & 19$\pm$9    & 29$\pm$18\\
\hline
\bf{Gas} &  &75$\pm$25\tablenote{Dwek et al. \cite{dw97a}} & 385$\pm$12\tablenote{Andr\'e et al. \cite{an03}, average between samples A and C} & $\approx$ 0 & $\approx$ 0 &  $\approx$ 0\\
\hline
\bf{Dust}& Solar  & 170$\pm$40 & 72$\pm$57 &   34$\pm$8 & 32$\pm$3    & 28$\pm$3\\
 &  F \& G stars  & 283$\pm$86 & 60$\pm$156 &  43$\pm$17 & 40$\pm$13    & 28$\pm$8\\
& B stars  & 115$\pm$81 & 0$\pm$134 & 		    23$\pm$7 & 19$\pm$9    & 29$\pm$18\\
\hline
\bf{Models} & ZDA (BARE-GR-S)  & 246 & 133  & 33 & 33   & 33 \\
& Li \& Draine \cite{ld01}   & 254 & 192  & 48 &48    & 48 \\
& ZDA (COMP-NC-B )  & 196 & 154  & 28 & 28   & 28 \\
\hline
\end{tabular}
\caption{Inferred dust phase abundances in the diffuse ISM}
\label{tab:a}
\end{table}

\begin{figure}
    \includegraphics[height=3.0in]{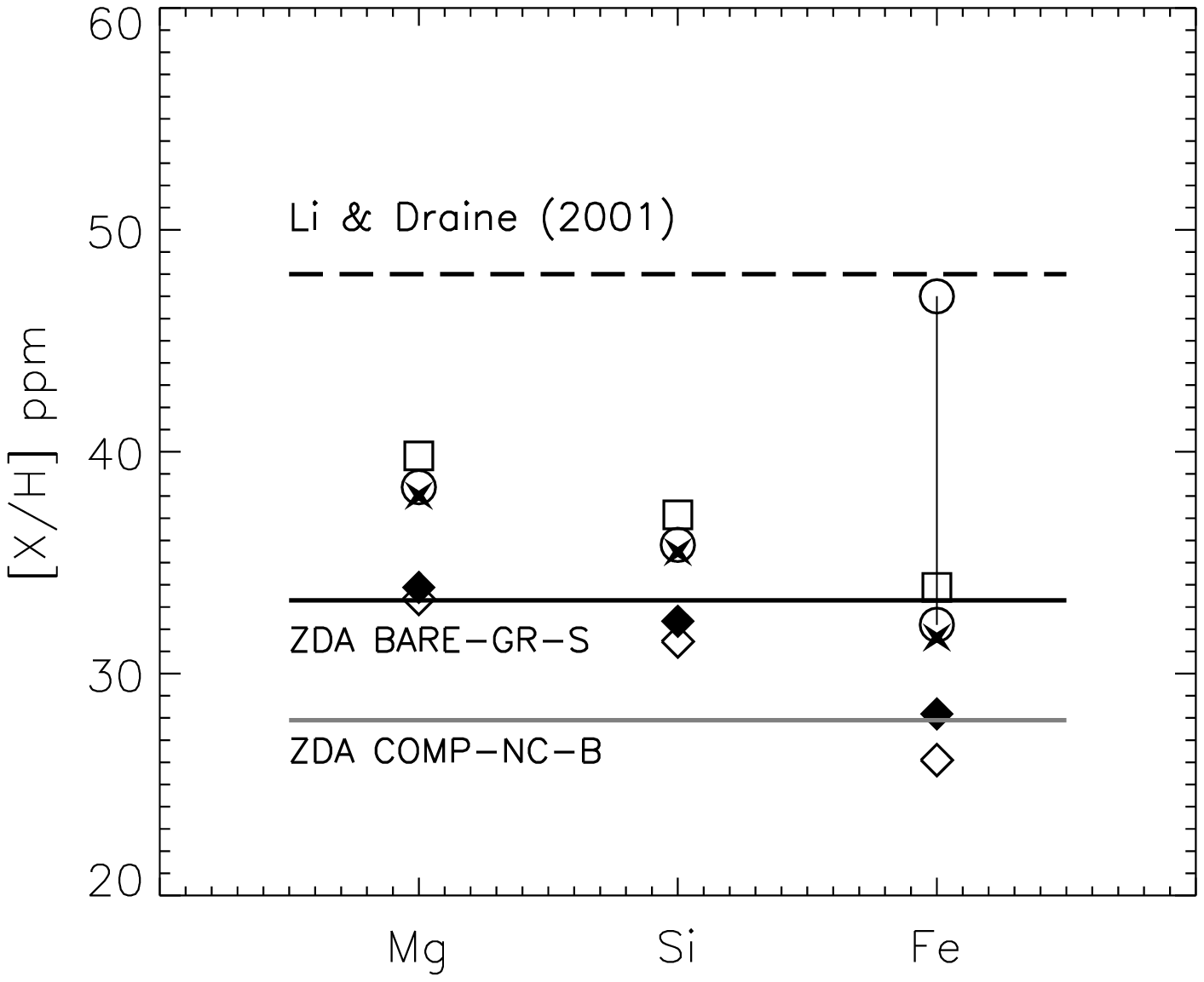}
  \caption{The relative steadiness of the solar (or meteoritic) abundance determinations of  Mg, Si, and Fe: Cameron \cite{c70}-{\it open diamond}; Gehren \cite{g88}-{\it open square}; Anders \& Grevesse \cite{ag89}-{\it open circle}; Grevesse \& Sauval \cite{gs98}-{\it filled star} ; Asplund, Grevesse, \& Sauval \cite{ags}-{\it filled diamond.} The Fe abundance of Anders \& Grevesse \cite{ag89} is represented by  both, the meteoritic and solar abundance determination because of the large discrepancy between the two values. The horizontal lines represent the results of the models discussed in the text.}
\end{figure}

The first ZDA model (BARE-GR-S) consists of bare silicate and graphite grains and PAHs, and was derived using the Holweger \cite{h01} solar abundances constraint. The dust composition and optical properties are identical to those of the Li-Draine model. However, they differ significantly in their grain size distribution (see Figure 19 and Table 7 in ZDA for the comparison and the analytical fit to the derived ZDA size distribution). Table~1 shows that the ZDA BARE model reproduces the updated abundances constraints better than the standard Li-Draine model, with the strictest constraint provided by the Mg, Si, and Fe abundance. The Li-Draine model requires $\sim$ 70\% more Fe and $\sim$ 50\% more Mg or Si to be in the dust than is available from either set of stellar abundances. Figure~1 shows how the solar abundance determinations of Mg, Si, and Fe varied over time, and compares them with the three dust models listed in the table. The abundance determination of Mg and Si have remained quite constant over time and consistent with meteoritic abundance determinations. The formerly large discrepancy between the solar and meteoritic Fe abundances reported by Anders \& Grevesse \cite{ag89} has been resolved by the more recent measurements of Asplund, Grevesse, \& Sauval \cite{ags}, which has settled on the lower value of 28$\pm$3.
Also listed in the table is model COMP-NC-B from ZDA, which consists of silicates, composite grains, and PAHs. All the carbon in this model is in PAHs and in the organic refractory  component of the composite grains. This model was constructed to fit the B star abundances and requires the least amount of carbon to be locked up in the solid phase of the ISM.

\subsection{Should there be a universal dust model?}

ZDA discovered a total of fifteen viable dust models that satisfy the extinction, IR emission and abundances constraint in the local ISM. Is there any way to discriminate between these models?
COMP grain models differ from BARE ones, because a significant fraction of their dust particles are fluffy composites containing voids. Composite particles have therefore an effective electron density that is significantly smaller than that of the bare particles. Consequently, observations of X-ray halos can, in principle, discriminate between the different classes of viable dust models \cite{dw04a} since X-ray halos are primarily produced by X-rays scattering off  electrons in large grains. However, X-ray halos sample a very limited fraction of the general ISM. So even if an X-ray halo would favor one model,  it would not preclude the viability of others in different regions of the ISM, since dust properties exhibit significant variations along different lines of sight.  

Some of the observational evidence for such variations are:
\begin{itemize}
\item the variations in the steepness of the FUV rise and the strength of the 2175 \AA \ bump \cite{f04} 
\item  the richness of mineral structures and ices seen in the evolved stars that are absent in the diffuse ISM \cite{w04}
\item variations in the elemental depletion pattern  in the hot, warm, and cold phases of the ISM \cite{s04}
\item variations in the PAH features in different regions of the ISM \cite{b04, on04}
\item variations in the strength and width of the silicate features \cite{d03}
\end{itemize}

These variations probably result from the existence of a large variety of dust sources producing dust with different composition and mineral structure, and from the fact that the ISM is not homogeneously mixed. Furthermore, grain processing in the ISM by thermal sputtering, grain-grain collisions, grain coagulation, and accretion in clouds plays an important role in producing large spatial variations in dust properties \cite{j04}. Such variations are manifested in the observed UV-optical extinction and IR emission from galaxies. 
Evidence for spatial variations in the extinction was provided by Keel \& White \cite{kw01}, who analyzed the extinction properties of dust in spiral galaxies that are partially backlit by an elliptical one, and by Clayton \cite{c04}, who reviewed extinction studies of the Magellanic Clouds and other nearby galaxies. There are too many observations showing the spatial variations in the IR emission spectrum from galaxies to list here, but many can be found in the special issue of The Astrophysical Journal Supplement  Series (volume 154) reporting the first results from the {\it Spitzer} satellite. 

\section{The Evolution of Dust}
\subsection{Dust sources and relative contributions}
 It is clear that there is no universal dust model that can be applied to a galaxy as a whole, or to galaxies with different evolutionary histories. Consider the production of dust from an evolving single stellar population (SSP) with masses between $\sim$ 0.8 and 40~M$_{\odot}$. The first dust that will be injected into the ISM will probably be formed in late-type Wolf-Rayet (WR) stars. Dust formation has only been observed to occur in the coolest C-rich stars of type WC8 and WC9. At least in a few cases, the formation of dust in these objects was induced by the interaction of the WR ejecta with the wind from a companion O-star \cite{mtd}. WR stars are however minor sources of dust that overall contribute less than 1\% of the total mass of dust injected by supernovae and AGB stars into the ISM \cite{jt, dw98}. After about  5~Myr the first stars of the SSP will undergo core collapse giving rise to Type II supernova (SN) events.  SN ejecta contain layers that are C- and O-rich, which are not intermixed on a molecular level. Consequently, SNe can produce both carbon and silicate type dust particles \cite{khn}. Stars with masses below $\sim$ 8~M$_{\odot}$ will undergo the AGB phase, lose mass, and evolve into white dwarfs. Figure 2 depicts the carbon and silicate yields from AGB stars with an initial solar metallicity. Stellar yields were taken from Marigo \cite{m01}. Stars with a C/O~$>$~1 ratio in their ejecta were assumed to condense only carbon dust, whereas stars with a C/O~$<$~1 ratio were assumed to condense only silicate type dust. The yields were calculated assuming a condensation efficiency of unity in the ejecta. The mass range of carbon producing AGB stars is between $\sim$ 1.6 and 4~M$_{\odot}$, a range that widens at lower stellar metallicities \cite{dw98}. So carbon dust from AGB stars will first be injected into the ISM after about 200~Myr, when  $\sim$ 4~M$_{\odot}$ stars evolve off the main sequence. 
 
 The AGB yields depicted in Figure 2 represent an idealized situation. In reality, AGB stars undergo thermal pulsations (the explosive ignition of the He-rich shell), that cause the convective mixing of C-rich gas with the outer stellar envelope. After repeated thermal pulses a star can evolve from an O-rich giant to a C-rich star. The changing composition of the stellar envelope will affect the chemistry of dust formation. Detailed kinetic nucleation calculations (Ferrarotti \& Gail \cite{fg}) show that AGB stars of a given mass can indeed form both, carbonaceous and silicate, type dust particles. Including the effect of radiation pressure on the newly-formed dust on the dynamics of the envelope, they find AGB yields that are significantly lower, by factors between 3 and 10, from those presented in Figure 2. A similar conclusion was reached by Morgan \& Edmunds \cite{me03} using a simpler model for dust formation in AGB stars.
 
 Figure 3 shows the carbon and silicate yields from both, AGB stars and SN~II, weighted by the stellar initial mass function (IMF), taken to be the Salpeter IMF between 0.7 and 40~M$_{\odot}$. SNe yields were taken from Woosley \& Weaver \cite{ww95}, and a condensation efficiency of unity was adopted in calculating the dust yield. The figure shows that the main contributors to the interstellar carbon abundance are low mass AGB stars, whereas SN~II are the main contributors to the silicate abundance in  the ISM. We emphasize however, that the yields presented in the figure are ideal ones, and that the actual yield of dust in SNe and AGB stars is still highly uncertain (see \S4 below).
 
\begin{figure}
    \includegraphics[height=3.0in]{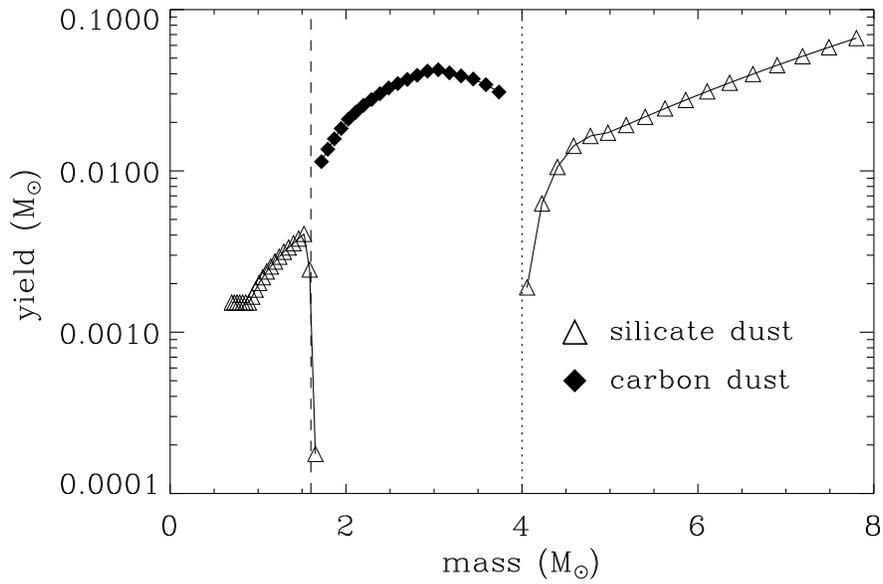}
  \caption{The carbon and silicate yield from AGB stars, based on the Marigo \cite{m01} yields.}
\end{figure}

\begin{figure}
    \includegraphics[height=3.0in]{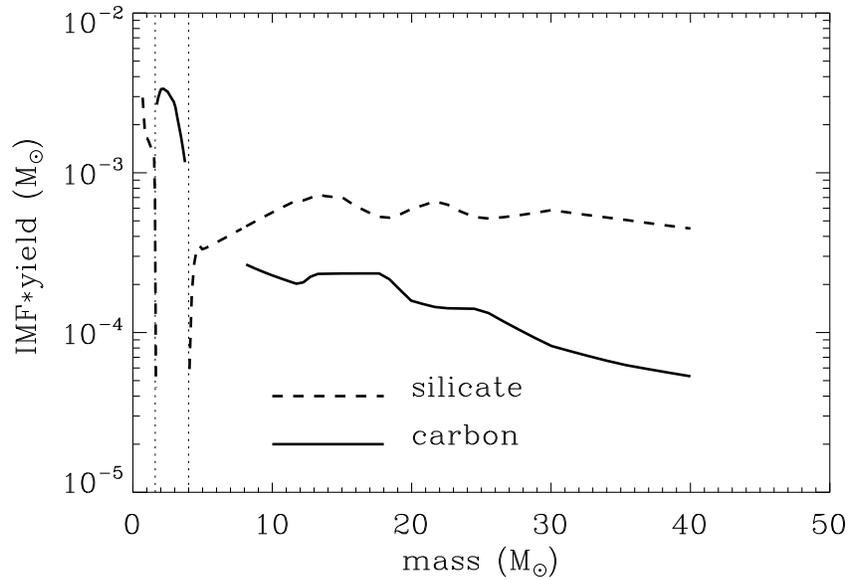}
  \caption{The IMF-weighted carbon and silicate yields from AGB stars and Type II supernovae. SN yields were taken from Woosley \& Weaver \cite{ww95}.}
\end{figure}

 \subsection{Dust processing in the ISM}
Following their injection into the ISM the newly-formed dust particles are subjected to a variety of interstellar processes, that result in the exchange of elements between the gas and dust phases of the ISM. These include:
 \begin{itemize}
\item thermal sputtering in high-velocity ($>$200 km~s$^{-1}$) shocks;
\item evaporation and shattering by grain-grain collisions in lower velocity shocks; and
\item accretion in dense molecular clouds.
\end{itemize}
Detailed description of the various grain destruction mechanisms and grain lifetimes in the ISM were presented by Jones, Hollenbach, \& Tielens \cite{jht} and recently reviewed by Jones \cite{j04}.

\subsection{The destruction of SN condensates by the reverse shock}
To these destruction processes we add the destruction of SN-condensed dust grains by the reverse shock propagating through the SN ejecta. The reverse shock is caused by the interaction of the ejecta with the ambient medium. Figure 4 is a schematic reproduction of a similar figure in \cite{tm99}, depicting the interaction of the SN ejecta during the free expansion phase of its evolution with its surrounding medium. This medium could consist of either circumstellar material that was ejected by the progenitor star during the red giant phase of its evolution, or interstellar material.  
The SN ejecta acts like a piston driving a blast wave into the ambient medium. Immediately behind the blast wave is a
region of shocked swept-up gas. When the pressure of this shocked gas exceeds that of the cooling
piston, a reverse shock will be driven into the ejecta \cite{m74}. 

\begin{figure}
    \includegraphics[height=3.0in]{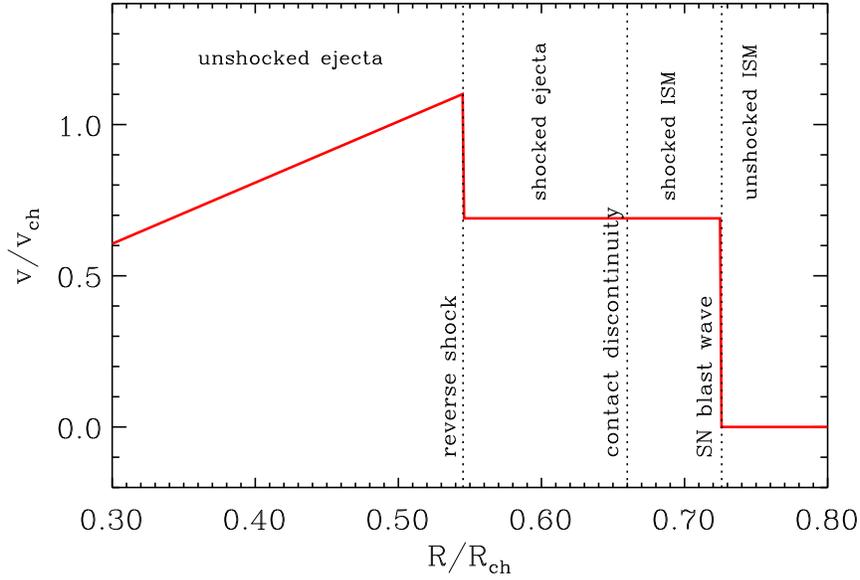}
  \caption{A schematic diagram (after Truelove \& McKee \cite{tm99}) depicting the interaction of the SN ejecta with its ambient surrounding.}
\end{figure}

Dust formed in the ejecta will be subject to thermal sputtering by the reverse shock. The fraction of dust destroyed is roughly given by the ratio of the sputtering lifetime, $\tau_{sput}$, to the expansion time (age), $t$, of the ejecta. The grain lifetime is initially a strongly rising function of gas temperature, reaching a plateau at about 10$^6$~K \cite{dw96}.  Figure 5 depicts the velocity history of the reverse shock as it traverses different layers of the ejecta, as a function of $\alpha \equiv R_r/R_{ej}$, where $R_r$ is  the radius of the reverse shock, and $R_{ej}$ is the outer radius of the ejecta. The calculations were performed using the analytical expressions of Truelove \& McKee \cite{tm99} for a SN explosion in a uniform medium.
The initial velocity of the reverse shock at $\alpha$ = 1 is zero, reaching a maximum at $\alpha$ = 0, when it reaches the origin of the explosion. No dust will be destroyed at $\alpha$ = 1, since the gas temperature so low that most gas molecules have kinetic energies well below the sputtering threshold. Very little grain destruction is also expected to take place at $\alpha$ = 0 since in spite of the high gas temperature, the gas density is very low and the sputtering lifetime is longer than the expansion time of the ejecta. There is therefore an optimal location 0 $< \alpha <$ 1, where the shock velocity (gas temperature) and ejecta density are such that $\tau_{sput}/t < 1$, and grain destruction can take place. 

\begin{figure}
    \includegraphics[height=3.0in]{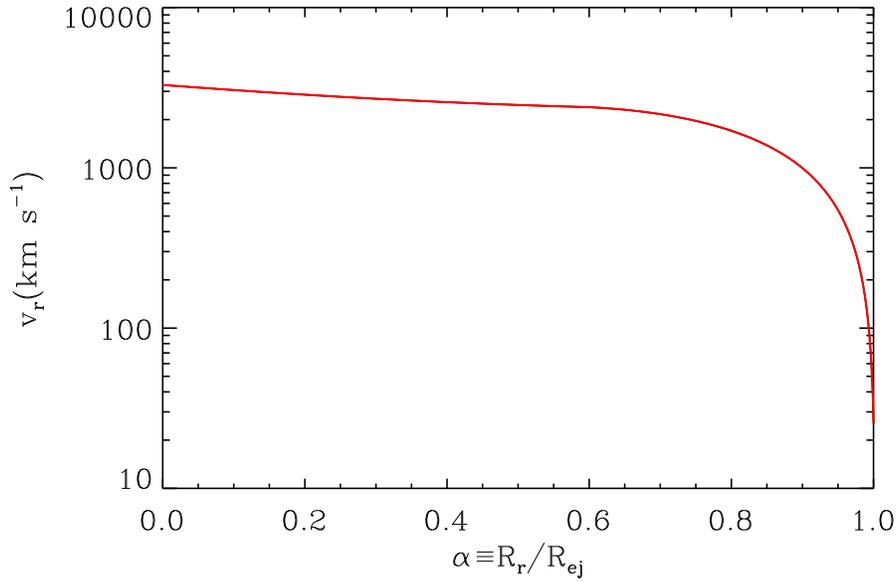}
  \caption{The velocity profile of the reverse shock traversing the SN ejecta. The reverse shock originates at $\alpha$ = 1, and over time propagates back into the ejecta, until it reaches the origin at $\alpha$ = 0.}
\end{figure}

\begin{figure}
    \includegraphics[height=3.0in]{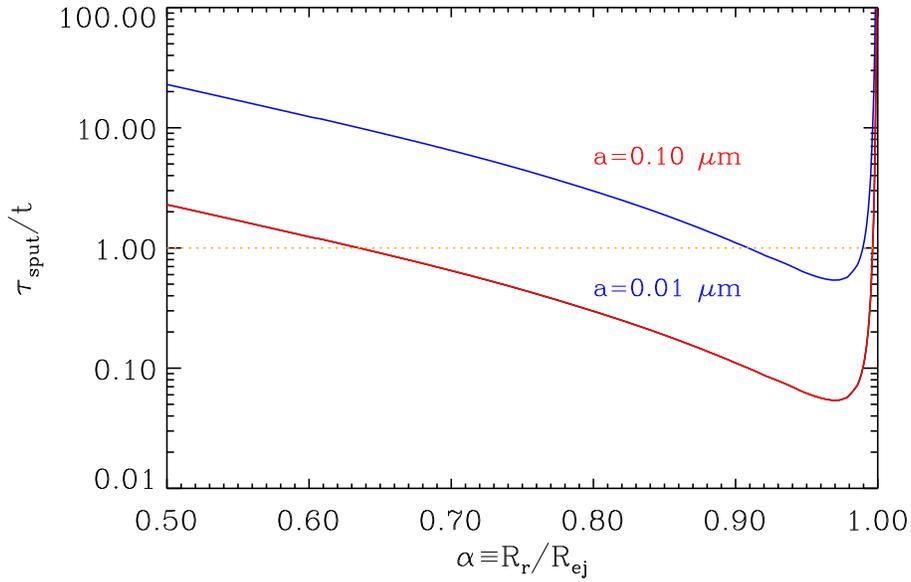}
  \caption{The survival of SN condensates in different layers of the ejecta. The survival of the dust is measured by the ratio of the sputtering timescale, $\tau_{sput}$ to that of the expansion time, $t$, of the ejecta. Grains are destroyed in layers for which $\tau_{sput}/t < 1$. }
\end{figure}

The $\alpha$-interval in which grains are completely destroyed will depend on the size of the newly-nucleated dust particles. Figure 6 depicts the location in the ejecta in which dust is completely destroyed by the reverse shock. The calculations were performed for dust particles with radii of 0.1 and 0.01 $\mu$m embedded in a smooth, O-rich ejecta. As expected, the smaller dust particles are destroyed over a wider range of ejecta layers compared to the larger size particles. In reality, SN ejecta are clumpy, and the SN dust is expected to reside predominantly in the clumpy phases of the ejecta, as is suggested by the detection of dust in the fast moving knots of the remnant of Cas~A \cite{l96, adm}.  The reverse shock slows down below the threshold for complete grain destruction as it traverses these density enhancements in the ejecta. Consequently, dust in the clumpy ejecta may only be shattered instead of being completely destroyed by sputtering. The total amount of grain processing in the  SN ejecta is however still highly uncertain. 
An independent investigation into the effect of reverse shocks from the H-envelope, the presupernova wind, and the ISM on the formation of dust, the amount of grain processing, and the implantation of isotopic anomalies in SN ejecta was carried out by Deneault, Clayton, \& Heger \cite{dch}.

 \subsection{Putting it all together in an idealized evolutionary model}

 Figure 7 depicts the evolution of the overall metallicity of the ISM (gas and dust), and that of the dust (silicates + carbon dust) in a normal galaxy with an exponential star formation rate characterized by a decay time of 6~Gyr. Starting with an initial star formation rate of 80~M$_{\odot}$~yr$^{-1}$, the galaxy will form about 3$\times 10^{11}$~M$_{\odot}$ of stars in a period of 13~Gyr. The silicate and carbon dust yields were calculated assuming a condensation efficiency of unity in the ejecta, and grain destruction was neglected. The model therefore represents an idealized case, in which grain production is maximized, and grain destruction processes are totally ignored.  Also shown in the figure are the separate contributions of AGB stars to the abundance of silicate and carbon dust. The onset of the AGB contribution to the silicate abundance starts when $\sim$ 8~M$_{\odot}$ stars evolve off the main sequence, whereas the AGB stars start to contribute to the carbon abundance only when  4~M$_{\odot}$ stars reach the AGB phase. The figure also presents the dust-to-ISM metallicity ratio. The ratio is almost constant at a value of $\sim$ 0.36. At $t$ = 14~Gyr, the model gives a silicate-to-gas mass ratio of 0.0048, and a carbon dust-to-gas mass ratio of 0.0025, in very good agreement to their values in the local ISM. 
 
\begin{figure}
    \includegraphics[height=3.0in]{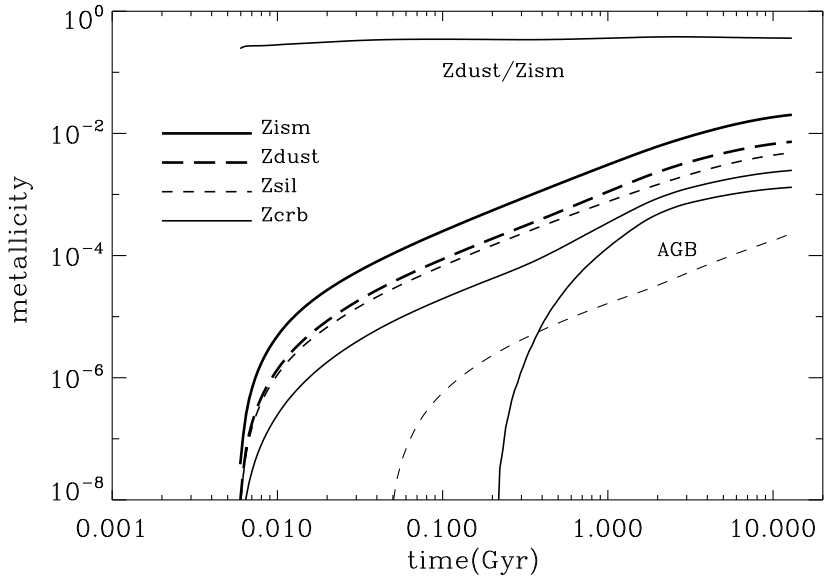}
  \caption{The evolution of the metallicity of the ISM and that of the dust as a function of time, Details in \S3.4 of the text.}
\end{figure}

\begin{figure}
    \includegraphics[height=3.0in]{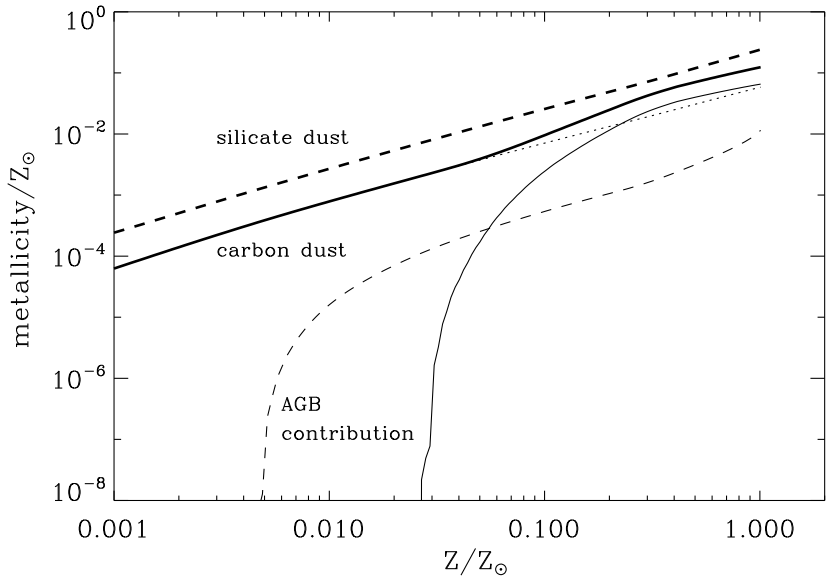}
  \caption{The evolution of dust as a function of ISM metallicity. Silicates are depicted by dashed lines, and carbon dust by solid lines. Bold lines represent the total contribution from SN~II and AGB stars, and the light lines the separate contributions of the latter sources.}
\end{figure}
 
 The idealized model presented above highlights several problems concerning the galactic evolution of dust:
\begin{enumerate}
\item the model reproduces the silicate and carbon dust abundances observed in the local ISM under idealized conditions. Any significant reduction in the yield of dust in SNe and AGB stars will result in a comparable reduction of the dust abundance; 
\item the above problem is exacerbated if grain destruction is taken into account, especially with the short timescales of 0.5 Gyr calculated by \cite{jht};
\item  an obvious solution is to postulate that the mass of interstellar dust is reconstituted by accretion onto surviving grains in molecular clouds. This solution poses a different set of problems, since the resulting morphology and composition of the dust may not be able to reproduce the constraints (extinction, IR continuum and broad emission features) observed in the local ISM \cite{n98};
\item however, the fact that ZDA discovered many dust models, including composite type particles that are expected to form in molecular clouds, that satisfy these observational constraint is a very encouraging solution to the interstellar dust abundance problem; 
\item finally, the global efficiency of grain destruction depends on the morphology of the interstellar medium, and may not be as high as calculated by \cite{jht}, especially if the filling factor of the hot cavities generated by expanding SN remnants is sufficiently large \cite{dw79}.
\end{enumerate}

  Figure 8 is a variation on the previous one, plotting select quantities as a function of the ISM metallicity. 
  The figure illustrates an interesting fact: if PAHs are only produced in AGB stars, then one would expect PAH features to arise in galaxies with a minimum metallicity of 0.1$Z_{\odot}$. This may be partly the cause for the very low abundance of PAHs in low metallicity systems \cite{g03, m04}, and for the appearance of PAH features in the spectra of galaxies only below a metallicity threshold of about  0.1$Z_{\odot}$ (Rieke and Engelbracht, private communications). 

The spectral appearance of a galaxy in the mid-IR and its UV-optical opacity is therefore affected by the delayed injection of carbon dust into the ISM. Figure 9 shows the evolution in the SED of a normal spiral galaxy as calculated by \cite{dw00}, illustrating the evolution of the PAH features with time. The heavy solid line represents the unattenuated stellar spectrum. The thin solid line is the total reradiated dust emission. At early epochs the reradiated IR emission is dominated by emission from H~II regions (top two panels), and therefore lack any PAH features. At later times, the contribution of non-ionizing photons dominates the dust heating, and consequently, the IR emission from the diffuse H~I gas dominates that from the H~II regions (dotted line, lower two panels). Also noticeable in the lower two panels is the difference between the attenuated and unattenuated stellar spectrum.

\begin{figure}
    \includegraphics[height=3.0in]{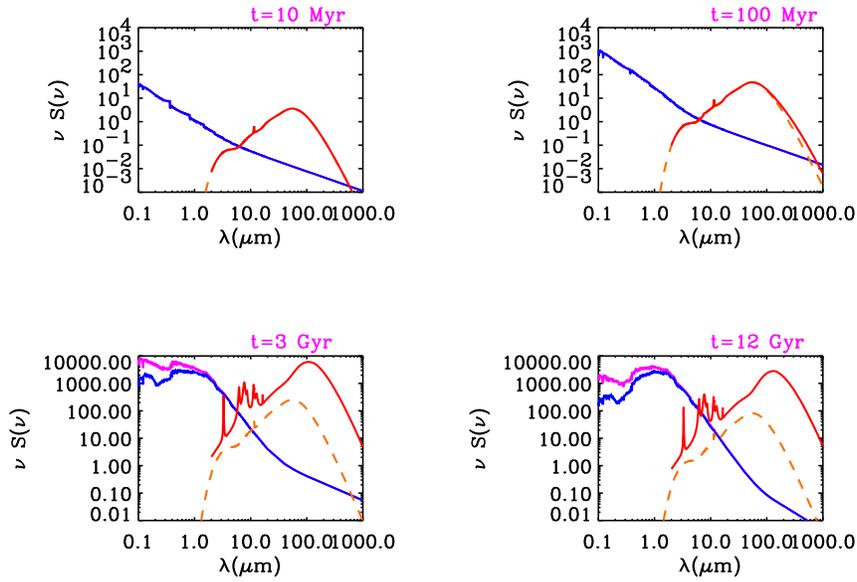}
  \caption{The evolution of the SED of a normal spiral galaxy as calculated by \cite{dw00}. More details in the text.}
\end{figure}

\section{When do galaxies become first opaque?}

On a cosmological scale, the formation and evolution of dust in galaxies and damped Ly$\alpha$ (DLA) systems has been a subject of considerable interest with the goals of studying the following: the effects of dust on the rate of various dust-related physical processes in their ISM such as the formation of H$_2$ \cite{hf}; the obscuration of quasars \cite{fp93, cne}; the relation between the dust abundance and galaxy metallicity \cite{lf, dw98}; the depletion of elements in DLA systems \cite{ks03, ks04}; and the evolution of the IR emission seen in the diffuse extragalactic background light \cite{pfh}. 

Also a subject of great interest is when galaxies became first opaque, opacity being defined in a bolometric sense as the fraction of total starlight energy that is processed by dust into IR emission.  Observations of ultraluminous IR galaxies at redshifts of $\sim$ 3 \cite{e04, i04} and the detection of large quantities of dust in high-redshift objects \cite{d03}, suggest that dust formation occurred early and efficiently after the onset of galaxy formation. 
The need for rapid dust formation has led Dunne et al. \cite{d03} to propose that massive, rapidly evolving stars must be responsible for the dust observed at high redshifts. This seems to be supported by the fact that dust production in AGB stars is delayed by a few hundred million years, compared to the production by SNe (see Figure 7). Furthermore, the yield of dust in AGB stars may be quite lower than that depicted in Figure 2. But what is the dust yield from SNe? Observations of Cas~A with the {\it ISO} \cite{l96, adm} and {\it Spitzer} satellites \cite{h04, k04} show that the total mass of SN condensed dust is less than $\sim$ 0.2~M$_{\odot}$, which is only about 10\%, of the total amount of refractory elements that formed in the ejecta. Using SCUBA submillimeter observations of the remnant, Dunne et al. \cite{d03} claimed to have detected a large mass ($M \approx 2 - 20$ M$_{\odot}$, depending on the adopted dust properties) of cold dust in the ejecta of Cas~A. This large amount of dust exceeds the amount of refractory elements produced in the explosion \cite{dw04b}. Subsequent observations with the {\it Spitzer} satellite, and comparison of the SCUBA data with molecular line observations revealed that the submillimeter emission from the direction of Cas~A is actually emitted from a molecular cloud along the line of sight to the remnant, instead of the ejecta \cite{k04, wb04}. The total mass of dust produced in SNe may therefore also be lower by a factor between 5 and 10 from those depicted in Figure 3. The question of what dust is responsible for the rise in galactic opacity will depend therefore on details of the dust evolution model. On one hand, SNe do indeed form the first dust, but on the other hand, carbon particles are significantly more opaque than silicates, potentially offsetting the advantage of SN condensates. What dust particles are ultimately responsible for producing the UV-optical opacity in  galaxies will therefore depend on the relative yields of SN- and AGB-condensed dust, and their subsequent evolution. Here we will only present a very preliminary investigation into this issue.

We will assume that galaxies first become opaque when the radial visual optical depth of molecular clouds in which most of the star formation takes place exceeds unity. The radial opacity of a cloud at $V$ is given by:
\begin{equation}
\tau(V) = Z_d \left({3M_c \over 4\pi R_c^2}\right)\kappa_d(V)
\end{equation}
 where $Z_d$ is the dust-to-gas mass ratio, $M_c$ the mass of the cloud, $R_c$ its radius, and $\kappa_d$ is the mass absorption coefficient of the dust. The criteria we adopt here is obviously a simplified one, since many overlapping optically thin molecular clouds can create an effectively  opaque line-of-sight to star forming regions. Nevertheless, this criterion is useful for this simple analysis. 
 Numerically, the  expression for $\tau(V)$ is approximately given by: 
 \begin{equation}
\tau(V) \approx 2\ Z_d \left({M_c/M_{\odot} \over \pi (R_c/{\rm pc})^2}\right) \left({\kappa_d(V)\over 10^4\ {\rm cm}^2\ {\rm g}^{-1}}\right)
\end{equation}

Dust opacities at $V$ are \cite{dl84}:
 \begin{eqnarray}
\kappa_d(V) & = & 3\times 10^3\ {\rm cm}^2 {\rm g}^{-1} \qquad \rm {for\ silicate\ dust} \\
  & = & 5\times 10^4\ {\rm cm}^2 {\rm g}^{-1} \qquad {\rm for\ carbon\ dust} 
\end{eqnarray}

 The surface density of molecular clouds in normal galaxies exhibits a narrow spread in values, ranging from about 10 to 100 M$_{\odot}$\ pc$^{-2}$ \cite{i00}. In luminous IR galaxies (LIRGs) star formation seems to take place in clouds with higher surface densities of about 10$^3$ to 10$^4$ M$_{\odot}$\ pc$^{-2}$ \cite{s92}. 
 
Adopting a cloud surface density of $5\times10^3$ M$_{\odot}$\ pc$^{-2}$ we get that actively star forming galaxies will become first opaque when $\tau(V) \approx $  1, or when
\begin{eqnarray}
Z_d & \approx & 3\times10^{-4}\qquad {\rm for\ silicate\ dust}  \\ \nonumber
& \approx & 2\times10^{-5}\qquad {\rm for\ carbon\ dust}
\end{eqnarray}

From Figure 7 we get that the critical metallicity for carbon dust is reached when the time lapse since the onset of star formation, $\Delta t$, is about 100~Myr, and that SN-condensed carbon dominates the abundance of carbon dust in the ISM. If only AGB stars produced carbon dust, then silicate dust particles will provide the first significant opacity and reach the critical metallicity at $\Delta t \approx$ 400~Myr. Carbon produced in AGB stars will reach the critical carbon metallicity later, at $\Delta t \approx$ 500~Myr. Adopting a standard $\Lambda$CDM cosmology with a Hubble constant of 70~km~s$^{-1}$~Mpc$^{-1}$, $\Omega_{\Lambda}$ = 0.73, and $\Omega_m$ = 0.27, we get that the rate at which the universe ages as a function of redshift $z$ is given, to an accuracy of $\sim$ 4\%, by:
\begin{equation}
\left|{dt\over dz}\right| = 7.4\ z^{-2}\ \ \ {\rm Gyr} \qquad {\rm for\ } 2 < z < 10
\end{equation}
If the SSP first formed at $z=z_s$, the the universe became first opaque at a redshift $z_{\tau=1}$ given by:
\begin{equation}
z_{\tau=1} \approx z_s\ \left[1+\left({\Delta t / {\rm Gyr} \over 7.4}\right)\ z_s\right]^{-1} \qquad 2 < z_{\tau=1} < z_s < 10
\end{equation}

\begin{figure}
    \includegraphics[height=3.0in]{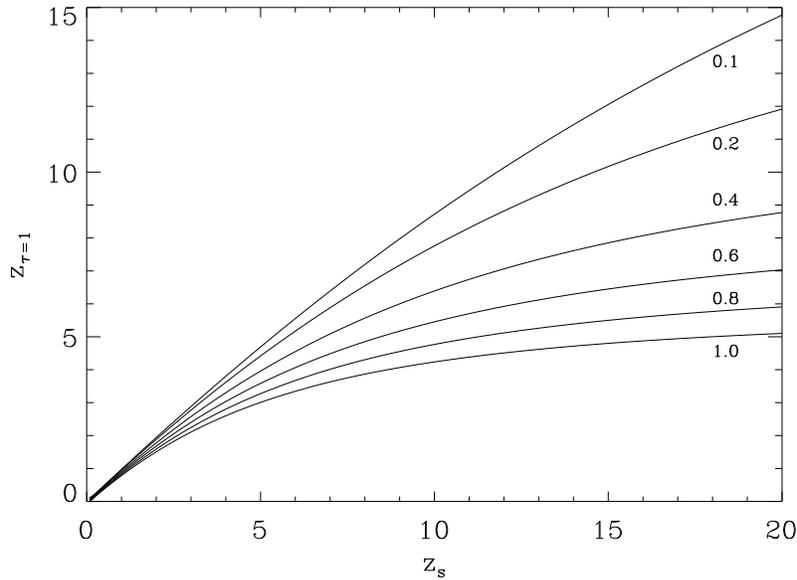}
  \caption{The redshift $z_{\tau=1}$, at which a galaxy becomes opaque (as defined in the text) as a function of the redshift $z_s$ at which the galaxy first formed. The different curves are labeled by $\Delta t$(Gyr), the time required for the dust abundance to be sufficiently high to cause typical molecular clouds to become opaque.}
\end{figure}

Figure 10 shows the exact relation between $z_s$ and $z_{\tau=1}$ for different values of $\Delta t$. For the simple model adopted here, the figure shows that a galaxy formed at redshift $z_s$ = 10, will become opaque at $z_{\tau=1} \approx$ 8.8 if $\Delta t$ = 100~Myr, and at $z_{\tau=1} \approx$ 5.9 if $\Delta t$ = 500~Myr. The actual value of $\Delta t$ depends on the chemical evolution model, the condensation efficiencies of carbon and silicate dust in the different sources, and the star formation history of the galaxy.

The figure presented here is very general, and illustrates the interrelation between the epoch of galaxy formation and the evolution of dust. It is applied here to a simplified model which was not specifically designed to follow the evolution of the ultraluminous IR galaxies (ULIRGs) observed at high redshifts. ULIRGs may have a much higher star formation rate than value of 80~M$_{\odot}$~yr$^{-1}$ used in the calculations. Furthermore, as mentioned before, the overlap of molecular clouds can render a galaxy opaque even when the individual clouds are optically  thin. The model above was only presented here for illustrative purposes, and can easily be used to solve the inverse problem:  given the fact that a galaxy is observed to be optically thick at a given redshift, what are the required star formation rate and dust formation efficiencies to make it optically thick at  that redshift?  

\section{Summary}
We have made many advances in our understanding of interstellar dust. However, many details about its origin and evolution are still unclear. Major unresolved issues are the efficiency of dust formation in the various sources, especially supernovae; the composition and the in-situ survival of the newly-formed dust; the efficiency of grain destruction in the ISM; the reconstitution of dust particles by accretion in molecular clouds and the resulting dust composition; and finally, the global effects of dust evolution on galactic opacities and the redistribution of stellar energies into infrared emission.

Currently operating and future space-, air-, and ground-based observatories will provide a wealth of new information which will go a long way towards addressing and solving many of the issues raised above.

\begin{theacknowledgments}
I thank Richard Tuffs and Cristina Popescu for organizing a scientifically stimulating conference in a beautiful surrounding, and Rick Arendt for suggesting a more accurate form for eq. (7) than appeared in the original version of the manuscript, and for comments that greatly enhanced the clarity of this paper. This work has been supported by the NASA OSS Long Term Space Astrophysics Program LTSA-2003-0065
 \end{theacknowledgments}



\begin{thebibliography}{99}
\bibitem{ag89} Anders E., Grevesse N., 1989, Geochim. et Cosmochim. Acta, 53, 197.
\bibitem{an03} Andr\'e M. K., et al., 2003, ApJ, 591, 1000.
\bibitem{adm} Arendt R. G., Dwek E., Moseley S. H., 1999, ApJ, 521, 234.
\bibitem{ags} Asplund M., Grevesse N., Sauval A. J., 2004, astro-ph/0410214.
\bibitem{b04} Bakes, E. L. O., Bauschlicher C. Jr., Tielens A., 2004, in "Astrophysics of Dust", eds. Witt A. N., Clayton G. C. , Draine B. T., ASP Conf. Series, vol. 309, ASP, San Francisco,  p. 731.
\bibitem{c70} Cameron A. G. W., 1970, Space Science Reviews, 15, 121.
\bibitem{c04} Clayton, G. C, 2004,  in "Astrophysics of Dust", eds. Witt A. N., Clayton G. C. , Draine B. T., ASP Conf. Series, vol. 309, ASP, San Francisco,  p. 57.
\bibitem{cne} Churches D. K., Nelson A. H., Edmunds M. G., 2004, MNRAS, 347, 1234.
\bibitem{dch} Deneault E. A.-N., Clayton D. D., Heger A., 2003, ApJ, 594, 312.
\bibitem{djj} d'Hendecourt L., Joblin C., Jones A., eds., 1999, "Solid Interstellar Matter: The ISO Revolution", Springer, Berlin.
\bibitem{dl84} Draine B. T., Lee, H. M., 1984, ApJ, 285, 89. 
\bibitem{d03} Draine, B. T., 2003, Ann. Rev. Astron. Astrophys. , 41, 241.
\bibitem{d04} Draine B. T., 2004, in "Astrophysics of Dust", eds. Witt A. N., Clayton G. C. , Draine B. T., ASP Conf. Series, vol. 309, ASP, San Francisco,  p. 691.
\bibitem{d03} Dunne L., Eales S., Ivison R., et al., 2003, Nature, 424, 285.
\bibitem{dw79} Dwek E., Scalo J. M., 1979, ApJ, 233, L81.
\bibitem{dw96} Dwek E., Foster S. M., Vancura O., 1996, ApJ, 457, 244.
\bibitem{dw97a} Dwek E., et al. 1997, ApJ, 475, 565.
\bibitem{dw97b} Dwek E., 1997, ApJ, 484, 779.
\bibitem{dw98} Dwek E., 1998, ApJ, 501, 643.
\bibitem{dw00} Dwek E., Fioc M., V\'arosi F., 2000, in "ISO Surveys of a Dusty Universe", eds. Lemke D., Stickel M., Wilke, Springer, Berlin, p.157.
\bibitem{dw04a} Dwek E., et al., 2004a,  in "Astrophysics of Dust", eds. Witt A. N., Clayton G. C. , Draine B. T., ASP Conf. Series, vol. 309, ASP, San Francisco,  p. 499.
\bibitem{dw04b} Dwek E., 2004b, ApJ, 607, 848
\bibitem{e04} Egami E., Dole H., Huang J.-S., et al., 2004, ApJS, 154, 130.
\bibitem{fp93} Fall S. M., Pei Y., 1993, ApJ, 402, 479.
\bibitem{fg} Ferrarotti A. S., Gail H.-P. 2004, preprint, submitted to A\&A. 
\bibitem{f04}Fitzpatrick E. L., 2004,  in "Astrophysics of Dust", eds. Witt A. N., Clayton G. C. , Draine B. T., ASP Conf. Series, vol. 309, ASP, San Francisco,  p. 33.
\bibitem{g03} Galliano F., Madden S., Jones A. P., et al., 2003, A\&A, 407, 159.
\bibitem{g88} Gehren T., 1988, Reviews in Modern Astronomy, 1, 52
\bibitem{gs98} Grevesse N., Sauval A. J., 1998, Space Sci. Rev., 85, 161.
\bibitem{h04} Hines D. C., Rieke G. H., Gordon K. D., et al., 2004, ApJS, 154, 290.
\bibitem{hf} Hirashita H., Ferrara A., 2002, MNRAS, 337, 921.
\bibitem{h01} Holweger H., 2001, in "Joint SOHO/ACE Workshop on Solar and Galactic Composition",  AIP Conf. Proc. 598, ed. Wimmer-Schweingruber R. F., AIP, New York, p. 23.
\bibitem{i00} Inoue A. K., Kamaya H., 2000, PASJ, 52, L47.
\bibitem{i04} Ivison R. J., 2005, this volume (astro-ph/0412084).
\bibitem{jt} Jones A. P., Tielens A. G. G. M. 1994, in "The Cold Universe", eds. Montmerle Th.,  Lada, Ch. J., Mirabel I. F., Tr\^an Thanh V\^an J., \'Editions Fronti\`eres, Gif-sur-Yvette. 
\bibitem{jht} Jones A. P., Tielens A. G. G. M., Hollenbach D. J. 1996, ApJ, 469, 740.
\bibitem{j04} Jones, A. P., 2004,  in "Astrophysics of Dust", eds. Witt A. N., Clayton G. C. , Draine B. T., ASP Conf. Series, vol. 309, ASP, San Francisco,  p. 347.
\bibitem{ks03} Kasimova E. R., Shchekinov Yu. A. 2003, Astrophys. Sp. Sci., 284, 433.
\bibitem{ks04} Kasimova E. R., 2004, in "The Spectral Energy Distribution of Gas Rich Galaxies: Confronting Models with Data", Heidelberg, 4-8 Oct. 2004, eds. C.C. Popescu \& R.J. Tuffs, AIP Conf. Ser., in press
\bibitem{kw01} Keel W. C., White III R. E., 2001, AJ, 122, 1369.
\bibitem{km96} Kim S.-H., Martin P. G., 1996, ApJ, 462, 296.
\bibitem{khn} Kozasa T., Hasegawa H., Nomoto K., 1989, ApJ, 344, 325.
\bibitem{k04} Krause O., Birkmann S. M., Rieke G. H., et al., 2004, Nature, 432, 596
\bibitem{k03} Kr\"ugel E. , 2003, "The Physics of Interstellar Dust", IOP Publishing, Bristol.
\bibitem{l96} Lagage P. O., Claret A., Ballet J., et al., 1996, A\&A, 315, L273.
\bibitem{ld01} Li A., Draine B. T. 2001, ApJ, 554, 778.
\bibitem{lf} Lisenfeld, U., Ferrara A., 1998, ApJ, 496, 145.
\bibitem{m04} Madden S. C., Galliano F., Jones A. P., Sauvage M. 2004, in "The Spectral Energy Distribution of Gas Rich Galaxies: Confronting Models with Data", Heidelberg, 4-8 Oct. 2004, eds. C.C. Popescu \& R.J. Tuffs, AIP Conf. Ser., in press
\bibitem{m01} Marigo P., 2001, A\&A, 370, 194.
\bibitem{mrn} Mathis J. S., Rumpl W., Nordsieck K. H., 1977, ApJ, 217, 425.
\bibitem{m76} Mathis J. S. 1996, ApJ, 472, 643.
\bibitem{m74} McKee C. F., 1974, ApJ, 188, 335.
\bibitem{mtd} Monnier J. D., Tuthill P. G., Danchi W. C. 2002, ApJ, 567, L137.
\bibitem{me03} Morgan H. L., Edmunds M. G. 2003, MNRAS, 343, 427.
\bibitem{n98} Nuth III J. A., Hallenbeck S. L., Rietmeijer F. J. M., 1998, Earth Moon \& Planets, 80, 73
\bibitem{on04} Onaka T., 2004,  in "Astrophysics of Dust", eds. Witt A. N., Clayton G. C. , Draine B. T., ASP Conf. Series, vol. 309, ASP, San Francisco,  p. 163.
\bibitem{pfh} Pei Y., Fall S. M., Hauser M. G., 1999, ApJ, 522, 604.
\bibitem{s92} Sakamoto K., Ishizuki S., Kawabe R., Ishiguro M., 1992, ApJ, 397, L27.
\bibitem{sw95} Snow T. P., Witt A. N. 1995, Science, 270, 1455.
\bibitem{sm01} Sofia U. J., Meyer D. M., 2001, ApJ, 554, L221.
\bibitem{s04} Sofia U. J., 2004,  in "Astrophysics of Dust", eds. Witt A. N., Clayton G. C. , Draine B. T., ASP Conf. Series, vol. 309, ASP, San Francisco,  p. 393.
\bibitem{tm99} Truelove J. K., McKee C. F., 1999, ApJS, 120, 299.
\bibitem{w04} Waters L. B. F. M., 2004,  in "Astrophysics of Dust", eds. Witt A. N., Clayton G. C. , Draine B. T., ASP Conf. Series, vol. 309, ASP, San Francisco,  p. 229.
\bibitem{w03} Whittet D. C. B. 2003, "Dust in the Galactic Environment", 2nd edition, IOP Publishing, Bristol.
\bibitem{wb04} Wilson T. L., Batrla W., 2004, A\&A, submitted, astro-ph/0412533.
\bibitem{wcd} Witt A. N., Clayton G. C., Draine B. T., eds., 2004, " Astrophysics of Dust", AIP Conf. Series vol. 309, ASP, San Francisco.
\bibitem{ww95} Woosley S. E., Weaver T. A., 1995, ApJS, 101, 181 
\bibitem{zda} Zubko V., Dwek E., Arendt R. G. 2004, ApJS, 152, 211. 

\end{thebibliography}
\end{document}